\documentclass[conference]{IEEEtran}
\usepackage{cite}
\usepackage{amsmath,amssymb,amsfonts}
\usepackage{algorithmic}
\usepackage{textcomp}
\usepackage{xcolor}
\usepackage{subcaption}
\usepackage{graphicx}
\usepackage{tikz}
\usetikzlibrary{svg.path}
\usepackage{scalerel}
\usepackage{booktabs}
\usepackage{array}
\usepackage{amsmath,amssymb,amsthm}
\newtheorem{definition}{Definition}
\newtheorem{lemma}{Lemma}
\newtheorem{proposition}{Proposition}
\newtheorem{theorem}{Theorem}

\usepackage[hyperfootnotes=false,hidelinks]{hyperref}

\newcommand{\arxiv}[1]{#1}

\newcommand{\acronym}{RELIQ}

\newcommand{\evNT}{69}

\newcommand{\evMedCombinedT}{64.8}

\newcommand{\evBelowLBT}{15}

\newcommand{\evNRz}{75}

\newcommand{\evMedCombinedRz}{76.1}

\newcommand{\evBelowLBRz}{53}

\newcommand{\evGmRestructureT}{14.6}
\newcommand{\evGmListT}{69.6}
\newcommand{\evGmCombinedT}{76.3}
\newcommand{\evGmRestructureRz}{44.1}
\newcommand{\evGmListRz}{56.8}
\newcommand{\evGmCombinedRz}{79.1}
\newcommand{\evNGainFiveT}{64}
\newcommand{\evNGainFiveRz}{65}
\newcommand{\evMaxCombinedBase}{211}
\newcommand{\evMaxCombinedDepth}{1}

\newcommand{\evNBumpSix}{37}

\newcommand{\evMaxCombined}{99.5}
\newcommand{\evMaxCombinedName}{ising\_n420}
\newcommand{\evSpearmanDensity}{-0.6}
\newcommand{\evSpearmanQubits}{0.6}
\newcommand{\evSpearmanPprs}{0.4}

\newcommand{\evMaxCombinedList}{2}

\newcommand{\evGmCombinedSmallT}{44.5}
\newcommand{\evGmCombinedLargeT}{89.3}
\newcommand{\evGmPassesZeroT}{69.6}

\newcommand{\evGmPassesMaxT}{76.3}
\newcommand{\evGmPortsTwoT}{76.3}
\newcommand{\evGmPortsMaxT}{71.7}
\newcommand{\evGmPortsMaxRestructureT}{0.6}
\newcommand{\evGmPortsMaxListT}{71.7}
\newcommand{\evGmWeightCombinedT}{12.2}

\newcommand{\evGmCombinedSmallRz}{51.7}
\newcommand{\evGmCombinedLargeRz}{88.8}
\newcommand{\evGmPassesZeroRz}{56.8}

\newcommand{\evGmPassesMaxRz}{79.1}
\newcommand{\evGmPortsTwoRz}{79.1}
\newcommand{\evGmPortsMaxRz}{51.2}
\newcommand{\evGmPortsMaxRestructureRz}{0.9}
\newcommand{\evGmPortsMaxListRz}{46.9}
\newcommand{\evGmWeightCombinedRz}{5.5}

\usepackage{tikz}
\usetikzlibrary{calc,positioning,arrows.meta,decorations.pathreplacing,fit,backgrounds}
\usepackage{amsmath,amssymb}

\definecolor{PauliX}{HTML}{F1735B}      
\definecolor{PauliY}{HTML}{B25388}      
\definecolor{PauliZ}{HTML}{130AA7}      
\colorlet{PauliXText}{black}
\colorlet{PauliYText}{white}
\colorlet{PauliZText}{white}
\colorlet{PauliIText}{black}

\definecolor{Overflow}{HTML}{C62828}    
\colorlet{Conflict}{Overflow}           
\definecolor{Select}{HTML}{238B45}      
\colorlet{LabelGrey}{black}             
\definecolor{Ancilla}{HTML}{FFD54F}     
\definecolor{AncillaEdge}{HTML}{B88A00} 
\colorlet{Merge}{black!55}              
\colorlet{RowShade}{Select!18}           
\colorlet{Band}{black!10}             
\colorlet{BandEdge}{black!55}           
\colorlet{BandLabel}{black}             
\definecolor{PanelShade}{HTML}{F2F7FD}  
\definecolor{PanelEdge}{HTML}{A9C4E8}   
\definecolor{CodeBox}{HTML}{EAF2FC}     
\definecolor{CodeBoxEdge}{HTML}{A9C4E8} 
\colorlet{PPMBackground}{white}         
\definecolor{MethodList}{HTML}{EDA100}
\definecolor{MethodRestructure}{HTML}{008300}
\definecolor{FlowStep}{HTML}{FCEFD0}    
\colorlet{FlowList}{FlowStep}
\colorlet{FlowRestructure}{FlowStep}
\colorlet{FlowResult}{black!8}

\newcommand{\cw}{0.46}   
\newcommand{\ch}{0.42}   

\tikzset{
  ppm cell/.style={minimum width=\cw cm, minimum height=\ch cm, inner sep=0pt, outer sep=0pt,
                   font=\footnotesize, anchor=center},
  ppm cell/X/.style={ppm cell, fill=PauliX, text=PauliXText, node contents={X}},
  ppm cell/Y/.style={ppm cell, fill=PauliY, text=PauliYText, node contents={Y}},
  ppm cell/Z/.style={ppm cell, fill=PauliZ, text=PauliZText, node contents={Z}},
  ppm cell/I/.style={ppm cell, text=PauliIText, node contents={I}},
  ppm cell/-/.style={ppm cell, node contents={}},
  ppm frame/.style={draw=black, line width=0.6pt},
  conflict/.style={draw=Conflict, line width=1.4pt},
  select/.style={draw=Select, line width=1.2pt},
  grey label/.style={font=\scriptsize, text=LabelGrey},
  panel title/.style={font=\small, anchor=west, inner sep=0pt},
  brace/.style={decorate, decoration={brace, amplitude=3pt, mirror}, line width=0.5pt},
  flow/.style={-{Stealth[length=4pt]}, line width=0.5pt},
  band/.style={fill=Band, draw=BandEdge, line width=0.5pt, rounded corners=1.5pt},
  band label/.style={font=\scriptsize, text=BandLabel, inner sep=1pt},
  metric/.style={font=\footnotesize, inner sep=1pt},   
}
\newcommand{\chres}{0.21}   

\newcommand{\cellbox}[4][conflict]{%
  \draw[#1] ({#2 - \cw/2}, {#3 - (#4-1)*\ch + \ch/2}) rectangle ({#2 + \cw/2}, {#3 - (#4-1)*\ch - \ch/2});
}
\newcommand{\overflowgauge}[2]{%
  \begin{scope}[shift={#1}, x=1cm, y=1cm]
    \filldraw[fill=white, draw=Overflow, line width=0.6pt] (-#2, 0) arc[start angle=180, end angle=0, radius=#2] -- cycle;
    \fill[Overflow] (0, 0) -- (#2*0.92, 0) arc[start angle=0, end angle=45, radius=#2*0.92] -- cycle;
    \fill[white] (#2*0.55, 0) arc[start angle=0, end angle=180, radius=#2*0.55] -- cycle;
    \foreach \a in {150, 120, 90} \draw[Overflow, line width=0.4pt] ({#2*0.62*cos(\a)}, {#2*0.62*sin(\a)}) -- ({#2*0.88*cos(\a)}, {#2*0.88*sin(\a)});
    \draw[Overflow, line width=0.9pt, line cap=round] (0, 0) -- ({#2*0.85*cos(12)}, {#2*0.85*sin(12)});
    \fill[Overflow] (0, 0) circle[radius=#2*0.14];
  \end{scope}%
}
\newcommand{\colbox}[4][select]{%
  \draw[#1] ({#2 - \cw/2 - 0.05}, {#3 + \ch/2 + 0.05}) rectangle ({#2 + \cw/2 + 0.05}, {#3 - (#4-1)*\ch - \ch/2 - 0.05});
}

\def\BibTeX{{\rm B\kern-.05em{\sc i\kern-.025em b}\kern-.08em
    T\kern-.1667em\lower.7ex\hbox{E}\kern-.125emX}}
\begin{document}

\definecolor{orcidlogocol}{HTML}{A6CE39}
\tikzset{
  orcidlogo/.pic={
    \fill[orcidlogocol]
    svg{M256,128c0,70.7-57.3,128-128,128C57.3,256,0,198.7,0,128C0,57.3,57.3,0,128,0C198.7,0,256,57.3,256,128z};
    \fill[white] svg{M86.3,186.2H70.9V79.1h15.4v48.4V186.2z}
    svg{M108.9,79.1h41.6c39.6,0,57,28.3,57,53.6c0,27.5-21.5,53.6-56.8,53.6h-41.8V79.1z
    M124.3,172.4h24.5c34.9,0,42.9-26.5,42.9-39.7c0-21.5-13.7-39.7-43.7-39.7h-23.7V172.4z}
    svg{M88.7,56.8c0,5.5-4.5,10.1-10.1,10.1c-5.6,0-10.1-4.6-10.1-10.1c0-5.6,4.5-10.1,10.1-10.1C84.2,46.7,88.7,51.3,88.7,56.8z};
  }
}

\newcommand\orcidicon[1]{\textsuperscript{\href{https://orcid.org/#1}{\mbox{\scalerel*{
          \begin{tikzpicture}[yscale=-1,transform shape]
            \pic{orcidlogo};
          \end{tikzpicture}
}{|}}}}}

\title{\texorpdfstring{\acronym{}}{RELIQ}: Optimizing Parallel Execution of Commuting Pauli Product Rotations
}

\arxiv{
\author{\IEEEauthorblockN{
Sayam Sethi\IEEEauthorrefmark{2}\IEEEauthorrefmark{1}\orcidicon{0009-0005-3056-5285},
Devika Nambisan\IEEEauthorrefmark{2},
Jonathan Mark Baker\IEEEauthorrefmark{2}\orcidicon{0000-0002-0775-8274}
}
\IEEEauthorblockA{
\IEEEauthorrefmark{2}Electrical and Computer Engineering, The University of Texas at Austin\\
\IEEEauthorrefmark{1}\href{mailto:sayams@utexas.edu}{sayams@utexas.edu}
}
}
}

\maketitle

\begin{abstract}
Fault-Tolerant Quantum Computation (FTQC) permits parallel execution of mutually commuting Pauli Product Rotations (PPRs), but per-qubit access point/port limits (e.g. two X and two Z edges on the surface code) force commuting groups that exceed the budget to execute sequentially, inflating circuit depth. We show that minimizing this hardware-limited depth is NP-hard and propose two heuristics for reducing it: 1. port-aware list scheduling, which reorders commuting PPRs by placing each PPR in the earliest group that can still accommodate it, and 2. generator restructuring, which rewrites each group as an equivalent generating set with reduced per-qubit port pressure. On QASMBench circuits compiled to PPRs, we combine the two heuristics in \acronym{} (\underline{Re}structure\,+\,\underline{Li}st schedule \underline{Q}uantum circuits) and observe an average hardware-limited depth reduction of \evGmCombinedT\% and \evGmCombinedRz\% for Clifford\,+\,T and Clifford\,+\,Rz circuits, respectively, over a non-reordering baseline. These gains persist as the per-qubit port budget grows, with \acronym{} still reducing the depth by \evGmPortsMaxT\% and \evGmPortsMaxRz\% at $16$ ports, suggesting these heuristics remain relevant even as hardware exposes more access points.
\end{abstract}

\arxiv{
}
\section{Introduction}

\begin{figure}
\centering
\includegraphics[width=\linewidth]{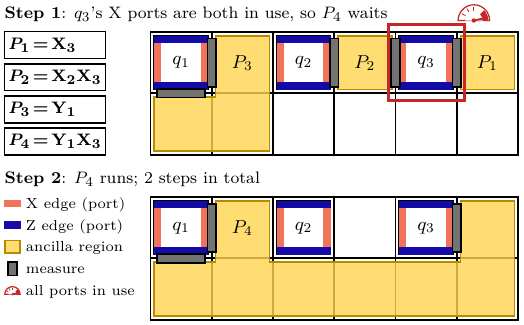}
\caption{\textbf{Port limits delay commuting Pauli Product Measurements (PPMs).} Each qubit patch exposes two $X$ (orange) and two $Z$ (blue) edges/ports, and each edge participates in at most one PPM. Ancilla regions are shaded in yellow. In step 1, $P_1$, $P_2$ and $P_3$ run in parallel, but $P_1$ and $P_2$ occupy both $X$ edges of $q_3$, so $P_4$ cannot start. Therefore, this sequence requires two steps despite all four measurements commuting with each other.}
\label{fig:device}
\end{figure}

As physical error rates drop below the thresholds of quantum error correction (QEC) codes (e.g., ${\sim}1\%$ for the surface code~\cite{fowler_surface_2012}), compilation tools tailored for error-corrected architectures become increasingly important. Prior work largely targets the physical layer~\cite{li_tackling_2019,khan_cyclone_2025,viszlai_matching_2025} or how a code implements logical operations~\cite{eastin_restrictions_2009,swaroop_universal_2024,ide_fault-tolerant_2025,williamson_low-overhead_2026}.
We consider the surface code equipped with lattice surgery~\cite{litinski_game_2018,horsman_surface_2012} and programs compiled to a sequence of Pauli Product Measurements (PPMs), although our techniques can easily be extended to other codes that support surgery via extractor-based systems~\cite{he_extractors_2025}. A set of PPMs can be executed simultaneously if they mutually commute~\cite{litinski_game_2018,cowtan_parallel_2026}, so without additional ancillas~\cite{fowler_time-optimal_2013,khan_architecting_2026}, the circuit duration is bounded by the longest chain of anti-commuting PPMs. However, hardware further restricts this parallelism. Each program qubit is assigned a single logical patch, whose $X$ and $Z$ edges act as ``ports'' to its logical operators, and each qubit can only participate in as many simultaneous PPMs as it has ports of the required type. For example, the four commuting PPMs in Figure~\ref{fig:device} need two steps, since three of them require an $X$ port on $q_3$, which has only two $X$ edges. This restriction can be lifted, but at the cost of non-trivial logical qubit overhead~\cite{cowtan_parallel_2026}.

In this work, we introduce an additional pass to the logical compilation pipeline that, to our knowledge, has not been explored before: maximizing the parallel execution of logical measurements under the limited access that the hardware provides to each logical qubit (Figure~\ref{fig:overview}). We show that minimizing this hardware-limited depth is NP-hard, and propose two heuristics: 1. Port-aware list scheduling reorders commuting PPMs, placing each in the earliest group that can still accommodate it, and 2. Restructuring replaces a mutually commuting group $\langle P_1, \ldots, P_s\rangle$ by an equivalent generating set with lower port loads and weight, e.g., replacing $P_i$ by $P_i\prod_{j\neq i} P_j^{\delta_j}$ with $\delta_j \in \{0, 1\}$. \acronym{} alternates both heuristics over several passes and reduces the hardware-limited depth on QASMBench by \evGmCombinedT\% and \evGmCombinedRz\% on average for Clifford\,+\,T and Clifford\,+\,Rz circuits, respectively.

\begin{figure*}
\centering
\input{figures/fig1_overview_body}
\caption{\textbf{Overview of \acronym{}.} (a)~Each PPR is transformed into a PPM with a $Z$ measurement on its own resource state. (b)~Mutually commuting PPMs form groups (ideal depth 2). (c)~The baseline greedy grouping of the input order pushes $P_6$ to a third step, since three $Z$ measurements on $q_2$ exceed the port limit, $c=2$. (d)~List scheduling moves $P_4$ before $P_3$, and Restructuring transforms $P_5'' \gets P_5'P_6'$, reducing the port load on $q_2$. (e)~\acronym{} (Section~\ref{sec:reliq}) starts from the List schedule $S_0$ (Section~\ref{sec:list}). Each subsequent pass $l=1, \ldots, \ell$ flattens the current schedule, restructures each commuting group (Section~\ref{sec:restructure}), and list-schedules the result, picking the schedule that minimizes (depth, weight).}
\label{fig:overview}
\end{figure*}
\section{Background}
\subsection{Quantum Error Correction}
An $[[n, k, d]]$ Quantum Error Correcting (QEC) code protects $k$ logical qubits against noise by encoding them into $n$ physical qubits, with a distance $d$ that sets how many physical errors can be corrected. 
The choice of QEC code determines its Pauli operators. For topological codes, e.g., the rotated surface code, these manifest as geometric boundaries of the patches, shown as orange ($X$) and blue ($Z$) edges in Figure~\ref{fig:device}. 
Therefore, the surface code only has 2 $X$ and 2 $Z$ edges available for measurement. Other codes can have different numbers of measurable edges available depending on the code properties and device connectivities~\cite{cowtan_parallel_2026,ismail_fast_2026}.

\subsection{Pauli-Based Computing (PBC)}
The surface code admits fault-tolerant Clifford gates and is made universal most commonly by producing $T$ states via magic state distillation factories~\cite{litinski_game_2018}. As such, circuits are often compiled into the Clifford\,+\,T gate set. The produced magic states are injected via gate teleportation.
An alternative to the logical gate-based operation is to instead perform all operations via projective Pauli measurements. Every Clifford\,+\,T circuit can be converted into a PBC circuit by commuting every Clifford operation to the end. All Clifford operations are absorbed into final measurements and the circuit becomes partitioned into two sets of operations: non-Clifford Pauli Product Rotations (PPRs) and Pauli Product Measurements (PPMs) (Figure~\ref{fig:overview}a). Each PPR consumes its own resource state and is executed as a PPM on the data qubits and that resource state (Section~\ref{sec:formulation}). The weight of the measurement is the number of non-identity Paulis in the measurement. 
The type of resource state depends on when this compilation pass occurs, e.g., if performed after compiling to Clifford\,+\,Rz, the resulting circuit is many $P(\theta)$ measurements which consume $R_P(\theta)$ states, whereas if performed after synthesizing all $Rz$ into $T$ then it consumes prepared $T$ states.


\subsection{Related Compilation Work}
Prior work addresses complementary layers of fault-tolerant compilation. Multiple prior works implement surface code routing and mapping under different assumptions~\cite{beverland_surface_2022,molavi_dependency_2025,silva_multiqubit_2024,leblond_resource_2024,sethi_rescq_2025}. Despite commuting gate sequences, these works execute them sequentially. Closest to our grouping problem, T-count and T-depth optimizers merge and parallelize commuting non-Clifford rotations~\cite{zhang_optimizing_2019,amy_polynomial-time_2013}, and measurement reduction for variational algorithms partitions commuting Pauli strings into simultaneously measurable families~\cite{gokhale_minimizing_2019,verteletskyi_measurement_2019}. Neither models per-qubit port limits, and \acronym{} can run on the output of T-count and T-depth optimizers. In contrast, code-surgery protocols enable parallel logical measurements through additional ancilla constructions~\cite{cowtan_parallel_2026,zhang_time-efficient_2025,yang_harnessing_2023}. \acronym{} instead targets the measurement representation and schedule under fixed per-qubit $X/Z$ port budgets without requiring additional ancilla constructions. \acronym{}'s optimization pass can be run in tandem with qubit mapping and routing protocols, and also benefits from additional ancilla constructions in the form of additional port availability to our optimizer.
\section{Optimizing Parallel Measurement Execution}\label{sec:solution}
We assume the input circuit $C = [P_1, \ldots, P_M]$ has already been compiled into PPMs by commuting the Cliffords and adding measurements on the resource states (Figure~\ref{fig:overview}a). Our approach works for both $T$ and $Rz(\theta)$ resource states and we ignore phases since they do not change commutations. Therefore, a group $S$ of mutually commuting PPRs can be performed in a single cycle if the following conditions hold:

\begin{align}
  & \label{eq:1}\lambda_{X,i}(S) = \sum_{P_j \in S} x_i(P_j) + y_i(P_j) \le c \quad i = 1, \ldots, N\\
  & \label{eq:2}\lambda_{Z,i}(S) = \sum_{P_j \in S} z_i(P_j) + y_i(P_j) \le c \quad i = 1, \ldots, N
\end{align}
where $p_i(P_j) = 1$ if the character $p$ occurs in the $i$-th position of the $P_j$ string and $0$ otherwise, $\lambda_{X,i}(S)$ and $\lambda_{Z,i}(S)$ are the $X$ and $Z$ port loads on qubit $i$, and $c$ is the number of $X$ (and $Z$) ports per logical qubit. For the surface code, $c = 2$ (Figure~\ref{fig:device}). We call such a group $S$ \emph{feasible}. Note that a $Y$-type port requires access to both the $X$ and $Z$ ports simultaneously. While we are hardware-aware when considering ports, we will be agnostic to the ancilla requirements needed to realize all of the simultaneous connections since this is handled separately by the mapping and routing compilation pass. 

\subsection{Problem Formulation}\label{sec:formulation}

Since every PPR $P_j$ must consume its own freshly prepared resource state $r_j$ (Figure~\ref{fig:overview}a), we denote by $P'_j = P_j \otimes Z_{r_j}$ the resultant PPM that includes the $Z$ measurement on its resource state, and by $C' = [P'_1, \ldots, P'_M]$ the input circuit as a sequence of PPMs. We treat the resource states as data qubits: since they are measured in the $Z$ basis, they do not change commutation, i.e., $P'_j$ and $P'_{j'}$ commute if and only if $P_j$ and $P_{j'}$ do, but they count toward the port loads and budgets in~\eqref{eq:1},~\eqref{eq:2}. Initially, every resource state starts with a port load of $1$. However, this can change after restructuring (Section~\ref{sec:restructure}). Two adjacent PPMs can be swapped if and only if they commute. To capture all such reorderings, we construct a DAG on the PPMs, $\mathcal{T} = (V, E)$ where $V = \{1, \ldots, M\}$ and we add an edge $(j, j') \in E$ for every $j < j'$ when $P'_j$ anti-commutes with $P'_{j'}$. Therefore, valid orders of $C'$ that can be generated by swapping commuting PPMs are exactly the topological sorts of $\mathcal{T}$~\cite{diekert1995traces}. We now formulate the scheduling problem as follows:

\begin{definition}[Port-constrained scheduling]\label{def:pcs}
A \emph{schedule} of depth $D$ is an ordered partition $[G_1, \ldots, G_D]$ of $C'$ such that (i) for every $(j, j') \in E$, $P'_j$ lies in an earlier group than $P'_{j'}$, and (ii) every $G_g$ is \emph{feasible}. The hardware-limited depth $D^\star(C')$ is the minimum such $D$. In the \emph{fixed-order} variant, the groups must also split $C'$ into consecutive blocks.
\end{definition}

Note that a schedule simply reorders $C'$ using commuting swaps and then groups the result, so $D^\star(C')$ is the best depth achievable by reordering commuting measurements. We can bound $D^\star(C')$ from below as follows:

\begin{proposition}[Lower bound]\label{prop:lb}
Let $h(\mathcal{T})$ be the number of vertices on a longest path of $\mathcal{T}$. Then
\begin{equation}
D^\star(C') \ge \max\left\{h(\mathcal{T}),\ \max_i\left(\left\lceil \frac{\lambda_{X,i}(C')}{c}\right\rceil,\left\lceil \frac{\lambda_{Z,i}(C')}{c}\right\rceil\right)\right\}.
\label{eq:lb}
\end{equation}
\end{proposition}

This bound holds for any reordering of $C'$. However, restructuring (Section~\ref{sec:restructure}) changes the port loads and therefore it is not bounded by~\eqref{eq:lb}. Unfortunately, computing $D^\star(C')$ exactly is intractable, even when every PPR has weight 2:

\begin{theorem}\label{thm:np}
For every $c \ge 1$, deciding whether $D^\star(C') \le 3$ is NP-complete, even if every $P_j$ has weight $2$.
\end{theorem}
\begin{proof}
A schedule can be checked in polynomial time. For hardness, we reduce from 3-edge-colouring of 3-regular graphs, which is NP-complete~\cite{holyer_np-completeness_1981}. Given such a graph $H = (U, F)$, each vertex $u \in U$ becomes a qubit with $c-1$ extra qubits $a_{u,1}, \ldots, a_{u,c-1}$, and each edge $\{u, v\} \in F$ becomes the ``edge'' PPM of the PPR $Z_uZ_v$. For every $u \in U$ and $r \in \{1, \ldots, c-1\}$, we also add the chain of PPMs of the PPRs $Z_uX_{a_{u,r}}$, $Z_uZ_{a_{u,r}}$, $Z_uX_{a_{u,r}}$. Only consecutive PPMs within a chain anti-commute, so in a depth-3 schedule the $g$-th PPM of every chain lies in $G_g$, and the chains of $u$ occupy $c-1$ of its $Z$ ports in every group. Hence the ``edge'' PPMs in each group form a matching, i.e., no two share a data qubit, so $G_1, G_2, G_3$ define a 3-edge-colouring. Conversely, placing the colour-$g$ edges of a 3-edge-colouring in $G_g$ gives a depth-3 schedule. Therefore, we get a reduction, proving NP-completeness.
\end{proof}

\begin{figure}
\centering
\includegraphics[width=\linewidth]{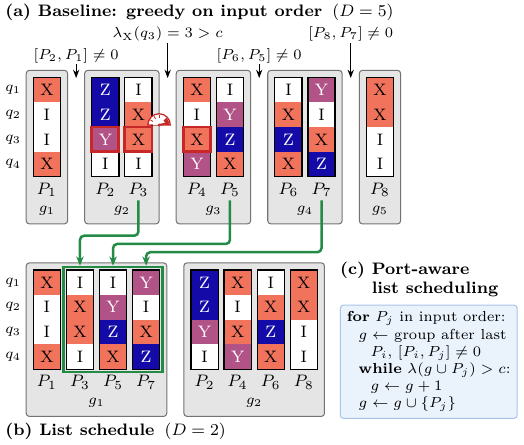}
\caption{\textbf{Port-aware list scheduling.} (a)~Greedy grouping in input order needs depth 5: three splits are anti-commutations, and $P_4$ commutes with $g_2$ but overflows $q_3$'s $X$ ports. (b)~$P_3$, $P_5$ and $P_7$ commute with every earlier PPR, and with $P_2$, $P_4$, $P_6$ only because two anti-commuting positions cancel, which gate-level dependency graphs miss. List scheduling moves them into $g_1$, giving depth 2. (c)~Each PPR joins the earliest feasible group after its last anti-commuting predecessor.}
\label{fig:list}
\end{figure}

Finally, constructing $\mathcal{T}$ requires checking commutation between every pair of PPMs, which runs in $O(M^2N/w)$ time using bitwise commutation checks on $w$-bit words. Given $\mathcal{T}$, the lower bound in~\eqref{eq:lb} can be computed with a longest-path pass in $O(M + |E|)$ and the port loads in $O(MN)$.

\subsection{Forming Mutually Commuting Groups}\label{sec:greedy}

Given the input circuit $C' = [P'_1, \ldots, P'_M]$, we can form the groups $[G_1, \ldots, G_D]$ in a straightforward greedy fashion. We start with an empty group $S$ and, for $j = 1, \ldots, M$, add $P'_j$ to $S$ if $S \cup \{P'_j\}$ is still \emph{feasible}, i.e., $P'_j$ commutes with every PPM in $S$ and no port load in~\eqref{eq:1},~\eqref{eq:2} exceeds $c$. Otherwise, we close $S$ as the next group $G_g$ and continue with $S = \{P'_j\}$. Without port limits ($c = \infty$), only anti-commutations split the groups, and the port limits can only split them further. We use this greedy grouping on the input order as our baseline, and it turns out to be optimal if the PPMs in the input circuit cannot be reordered:

\begin{lemma}\label{lem:greedy}
Greedy grouping is optimal if the order is fixed.
\end{lemma}
\begin{proof}
Any subset of a \emph{feasible} group is also \emph{feasible}, since subsets of commuting groups commute and port loads can only decrease. Let $a_g$ and $b_g$ be the number of PPMs in the first $g$ groups of the greedy grouping and of an optimal fixed-order schedule. By induction from $a_0 = b_0 = 0$, if $a_g \ge b_g$, then $P'_{a_g+1}, \ldots, P'_{b_{g+1}}$ lie in the optimal group $G_{g+1}$ and are thus \emph{feasible} together, so greedy group $g+1$ contains them and $a_{g+1} \ge b_{g+1}$. Hence the greedy grouping uses at most as many groups as the optimal fixed-order schedule.
\end{proof}

Therefore, any heuristic that improves over the baseline must either reorder (Section~\ref{sec:list}) or restructure (Section~\ref{sec:restructure}) the PPMs. Note that the greedy grouping runs in $O(MsN/w)$ time, where $s$ is the maximum group size.

\subsection{Port-Aware List Scheduling}\label{sec:list}

The greedy approach produces $D \ll M$ groups, i.e. the hardware-limited depth is substantially shortened by grouping into mutually commuting groups. However, this is non-optimal since the group boundaries depend on the order of the input circuit. Local permutations within a group can result in a re-grouping which has a lower hardware-limited depth. More formally, we can consider $C = [G_1, \ldots, G_D] = [\pi_1(G_1), \ldots, \pi_D(G_D)]$ where we have permuted elements within each group. We can re-stitch this back together into a new order and then again execute the greedy group formation. This may or may not produce a circuit with a lower hardware-limited depth. Every such order is a topological sort of $\mathcal{T}$ (Section~\ref{sec:formulation}), and therefore the circuit remains unchanged. However, finding the set of permutations $\{\pi_g\}$ which minimizes the hardware-limited depth is non-trivial as there are up to $\prod_g |G_g|! \le (s!)^D$ such permutations. When the average weight of PPRs is low (e.g. when every PPR is on average weight 2), the size of the groups is relatively large (since most PPMs will have disjoint supports on the data qubits and will thus commute) making this quickly intractable.

Therefore, instead of searching over permutations, we construct the schedule directly using port-aware list scheduling (Figure~\ref{fig:list}). We process the PPMs in the input order and place each $P'_j$ into the earliest group $G_g$ that (i) comes after the groups of all of its anti-commuting predecessors, i.e., of all $P'_{j'}$ with $(j', j) \in E$, and (ii) is still \emph{feasible} after adding $P'_j$. If no such group exists, we start a new group. By (i), $P'_j$ commutes with every PPM already in $G_g$, so only the port loads need to be checked. Unlike the greedy grouping, a PPM can therefore be placed into an earlier group, ahead of commuting PPMs that precede it in the input order. Moreover, list scheduling is never worse than the greedy grouping. It is also deterministic and requires a single pass over the PPMs. Finding the anti-commuting predecessors of every PPM requires constructing $\mathcal{T}$, which takes $O(M^2N/w)$ time, and each port check takes $O(N/w)$ time using bitsets of the saturated ports, so list scheduling runs in $O(M^2N/w)$ time.

\subsection{Restructuring the Groups}\label{sec:restructure}

To further minimize the hardware-limited depth, we need to minimize the port loads in~\eqref{eq:1},~\eqref{eq:2}. We first identify that the PPMs in a mutually commuting group are generators for a Pauli subgroup. Consider such a group $G_g$ obtained from the greedy grouping without port limits ($c = \infty$). The action of $G_g$ on the space is determined by the generators $\langle G_g \rangle$, which are independent since each acts on its own resource state. As such, we can construct an equivalent group by replacing these generators with any other generating set. Note that this also works for any potential fixup corrections needed since they all commute with each other as well, and the outcomes of the original PPMs can be recovered from the new ones.

\begin{figure}
\centering
\includegraphics[width=\linewidth]{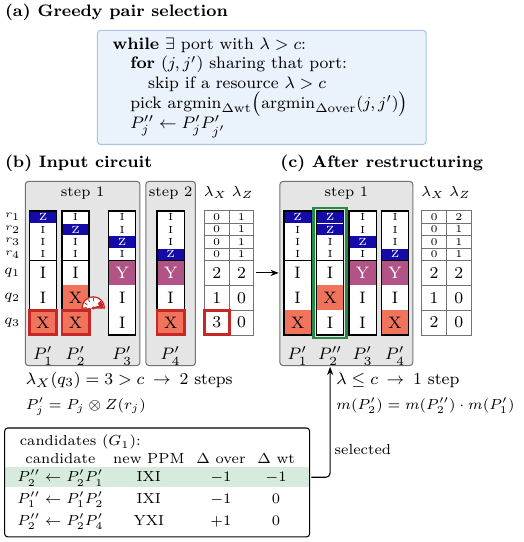}
\caption{\textbf{Restructuring commuting PPMs.} ($c=2$) (a)~Our greedy heuristic picks the PPM pair with the largest decrease in port load, breaking ties by weight reduction. PPMs remain candidates while their resource states stay within $c$, and the scan repeats until no pair reduces the overload. (b)~Four commuting PPMs, three of which need an $X$ port on $q_3$, require two steps. (c)~Replacing $P''_2\gets P'_2P'_1$ brings every load within $c$, so the group runs in one step. The original outcome is $m(P'_2)=m(P''_2)m(P'_1)$.}
\label{fig:restructure}
\end{figure}

For example, we can replace $P'_j$ with $P'_jP'_{j'}$ for any two $P'_j, P'_{j'}\in G_g$ and $j\neq j'$. This is beneficial when $P_j(i) = P_{j'}(i)$ for some qubit $i$, then the product has $P_jP_{j'}(i) = I$, thereby reducing the port load on $i$. However, the product always adds a $Z$ on the resource state of $P_{j'}$, and may \textit{increase} the port loads on other qubits as well.
Therefore, the objective is to select a new set of generators so as to simultaneously minimize the port loads on every qubit. The naive approach is to enumerate all such generators and select the one which minimizes the port loads in~\eqref{eq:1},~\eqref{eq:2}. However, this strategy is impractical since this requires enumerating all $2^{\Theta(s^2)}$ generating sets for each group $G_g$, where $|G_g| \leq s$.

We simplify the search by replacing one PPM at a time, i.e., replacing $P'_j$ by $P''_j = P'_jP'_{j'}$ and keeping all other PPMs identical as before. Each such replacement is invertible ($P'_j = P''_jP'_{j'}$), so the restructured group $G''_g$ generates the same group as $G_g$ and the original outcomes are recovered as $m(P'_j) = m(P''_j)\,m(P'_{j'})$. A PPM can take part in several replacements, but every replacement changes the resource states that $P''_j$ acts on, so we only allow a replacement if every resource state stays within its $c$ ports.
Even so, choosing the best sequence of replacements remains combinatorial. We therefore use a greedy heuristic (Figure~\ref{fig:restructure}a) that repeatedly applies the replacement with the largest decrease in the total port overload of the group, i.e., the total load in excess of $c$ over all qubits and resource states, breaking ties by the largest weight reduction, until no replacement decreases the overload. Since every replacement strictly lowers the overload, the heuristic stops after at most as many replacements as the initial overload. We also restrict the number of evaluated pairs to $10^8$ per pass.

\begin{figure*}
\centering
\includegraphics[width=\linewidth]{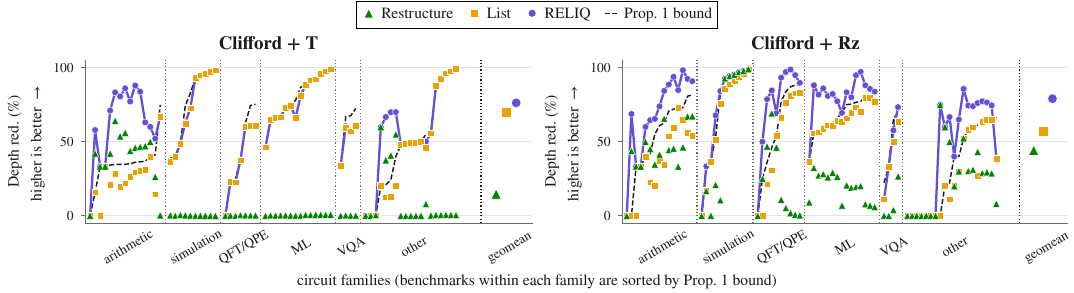}
\caption{\textbf{Depth reduction by benchmark family} at $c = 2$ for Clifford\,+\,T (left) and Clifford\,+\,Rz (right). Within each family, circuits are sorted by the bound of Proposition~\ref{prop:lb} (indicated by the dashed line). We also report the geometric-mean reduction.}
\label{fig:family}
\end{figure*}

\subsection{\texorpdfstring{\acronym{}}{RELIQ}: \texorpdfstring{\underline{Re}}{Re}structure\texorpdfstring{\,+\,}{ + }\texorpdfstring{\underline{Li}}{Li}st Schedule \texorpdfstring{\underline{Q}}{Q}uantum Circuits}\label{sec:reliq}
In \acronym{}, we combine restructuring with list scheduling. Starting from the list schedule ($l = 0$), each pass $l = 1, \ldots, \ell$ orders the PPMs by their group in the current schedule, restructures the mutually commuting groups, and list-schedules the result (Figure~\ref{fig:overview}e). Every pass replaces the generators invertibly, so the original outcomes can still be recovered. Over all passes, we select the schedule with the minimal hardware-limited depth, breaking ties by the total weight.
\begin{table}
\centering
\caption{\textbf{Comparisons over baseline by benchmark family} ($c = 2$, $\ell = 9$). Max: largest per-circuit reduction of \textsc{\acronym{}}. Weight: geometric-mean change (\%) of the total weight with \textsc{\acronym{}}. `vs Prop.~\ref{prop:lb} bound': \#circuits where \textsc{\acronym{}}'s depth is above ($>$), equal to ($=$) or below ($<$) the bound.}
\label{tab:qasmbench}
\scriptsize
\setlength{\tabcolsep}{0pt}
\begin{tabular*}{\columnwidth}{@{\extracolsep{\fill}}lrrrrrrw{c}{16pt}w{c}{16pt}w{c}{16pt}@{}}
\toprule
 & & & & \multicolumn{3}{c}{\textsc{\acronym{}}} & \multicolumn{3}{c}{vs Prop.~\ref{prop:lb} bound} \\
\cmidrule(lr){5-7}\cmidrule(l){8-10}
Family & $n$ & \textsc{List} & \textsc{Restr.} & geomean & max & Weight & $>$ & $=$ & $<$ \\
\midrule
\multicolumn{10}{@{}l}{\textbf{\emph{Clifford\,+\,T}}} \\
arithmetic & 15 & 27.3 & 41.3 & 68.9 & 88.0 & $-$40.8 & 1 & 3 & 11 \\
simulation & 10 & 87.2 & 0.2 & 87.2 & 98.3 & $-$0.1 & 3 & 7 & 0 \\
QFT/QPE & 7 & 41.8 & 0.2 & 41.9 & 61.0 & 0.0 & 4 & 3 & 0 \\
ML & 14 & 86.5 & 0.4 & 86.7 & 99.2 & $-$0.3 & 9 & 5 & 0 \\
VQA & 4 & 53.9 & 0.1 & 54.2 & 60.8 & $-$0.4 & 3 & 1 & 0 \\
other & 19 & 68.6 & 13.8 & 74.5 & 99.3 & $-$5.3 & 1 & 14 & 4 \\
\textbf{All} & 69 & 69.6 & 14.6 & \textcolor{Select}{\textbf{76.3}} & \textcolor{Select}{\textbf{99.3}} & \textcolor{Select}{\textbf{$-$12.2}} & 21 & 33 & \textcolor{Select}{\textbf{15}} \\
\midrule
\multicolumn{10}{@{}l}{\textbf{\emph{Clifford\,+\,Rz}}} \\
arithmetic & 15 & 40.9 & 45.7 & 80.0 & 98.3 & $-$18.1 & 0 & 3 & 12 \\
simulation & 11 & 83.8 & 85.5 & 89.9 & 99.5 & $-$0.1 & 0 & 3 & 8 \\
QFT/QPE & 10 & 59.8 & 25.3 & 88.0 & 98.7 & 1.3 & 0 & 1 & 9 \\
ML & 15 & 66.4 & 19.5 & 84.7 & 97.2 & $-$3.0 & 1 & 2 & 12 \\
VQA & 4 & 42.6 & 8.4 & 49.2 & 73.3 & $-$0.8 & 2 & 1 & 1 \\
other & 20 & 30.7 & 28.6 & 54.7 & 85.7 & $-$4.4 & 1 & 8 & 11 \\
\textbf{All} & 75 & 56.8 & 44.1 & \textcolor{Select}{\textbf{79.1}} & \textcolor{Select}{\textbf{99.5}} & \textcolor{Select}{\textbf{$-$5.5}} & 4 & 18 & \textcolor{Select}{\textbf{53}} \\
\bottomrule
\end{tabular*}

\end{table}

\begin{figure*}
\centering
\includegraphics[width=\linewidth]{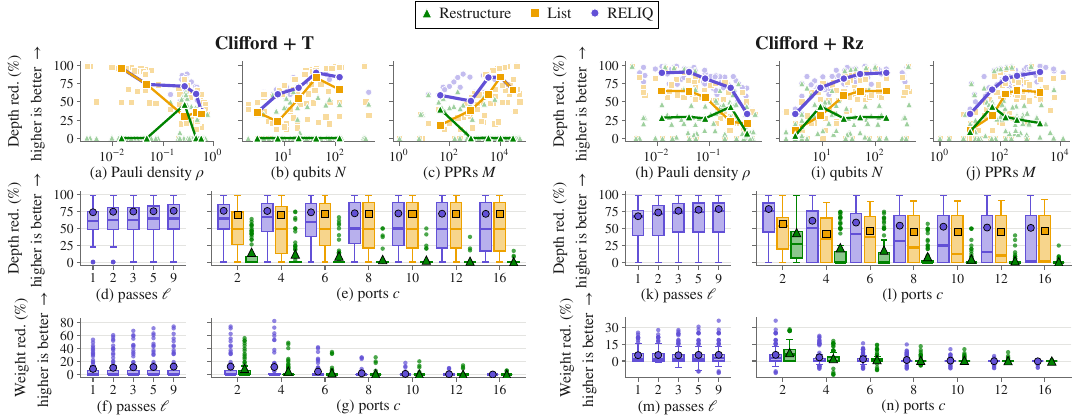}
\caption{\textbf{Sensitivity of the depth reduction.} (a)--(c), (h)--(j)~One point per circuit and method at $c = 2$ against Pauli density $\rho$, qubits $N$ and PPRs $M$, with lines showing medians of equal-count bins. (d), (k)~\textsc{\acronym{}} against passes $\ell$. (e), (l)~Depth reduction against ports $c$, versus the same-$c$ baseline. (f), (g), (m), (n)~Weight reduction against $\ell$ and $c$.}
\label{fig:sensitivity}
\end{figure*}

\section{Evaluation}\label{sec:eval}
\subsection{Setup}
We evaluate three strategies: \textsc{List} (Section~\ref{sec:list}), \textsc{Restructure} (Section~\ref{sec:restructure}), which restructures the groups of the input order before the greedy grouping, and \textsc{\acronym{}} (Section~\ref{sec:reliq}), which starts from the \textsc{List} schedule and picks the best of $\ell = 9$ passes of restructuring followed by list scheduling. All three strategies are deterministic. The baseline is the greedy grouping of the input order (Section~\ref{sec:greedy}), which by Lemma~\ref{lem:greedy} is the best schedule that maintains the input order. Therefore, every reduction comes from reordering, restructuring, or both. Unless stated otherwise, we use $c = 2$ and report the reduction in hardware-limited depth relative to the baseline at the same $c$.
Each input circuit is compiled to Clifford\,+\,T (via GridSynth~\cite{ross_optimal_2016}) and to Clifford\,+\,Rz, converted into PPRs by commuting the Cliffords through (Figure~\ref{fig:overview}a), and then into PPMs by adding measurements on the resource states (Section~\ref{sec:formulation}). We keep the \evNT{} Clifford\,+\,T and \evNRz{} Clifford\,+\,Rz circuits that contain non-Clifford rotations and compile within five minutes.

\subsection{Benchmarks}
We evaluate the strategies on the QASMBench suite~\cite{li_qasmbench_2022}. As Table~\ref{tab:qasmbench} shows, \textsc{\acronym{}} reduces the circuit depth by \evGmCombinedT\% and \evGmCombinedRz\% on average for Clifford\,+\,T and Clifford\,+\,Rz, respectively. Most of this gain comes from reordering, since \textsc{List} reduces the depth by \evGmListT\% and \evGmListRz\%, while \textsc{Restructure} reduces it by \evGmRestructureT\% and \evGmRestructureRz\%. \textsc{Restructure} has a smaller reduction for Clifford\,+\,T, since GridSynth makes consecutive rotations more likely to anti-commute. Despite its modest gains, \textsc{Restructure} in tandem with \textsc{List} in \textsc{\acronym{}} improves over the lower bound of Proposition~\ref{prop:lb} on \evBelowLBT{} Clifford\,+\,T and \evBelowLBRz{} Clifford\,+\,Rz circuits. \textsc{Restructure} also reduces the average PPM weight, while \textsc{List} cannot modify the weight since it only reorders the PPMs. The gains grow with the circuit size, from \evGmCombinedSmallT\% and \evGmCombinedSmallRz\% on small circuits to \evGmCombinedLargeT\% and \evGmCombinedLargeRz\% on large circuits. Figure~\ref{fig:family} shows the gains for every circuit, grouped by benchmark family. \evNGainFiveT{} of the \evNT{} Clifford\,+\,T and \evNGainFiveRz{} of the \evNRz{} Clifford\,+\,Rz circuits improve by at least $5\%$, with medians of \evMedCombinedT\% and \evMedCombinedRz\%, respectively. The largest gain, \evMaxCombined\%, is on \textsc{\evMaxCombinedName}, where the baseline needs \evMaxCombinedBase{} groups, \textsc{List} attains the Proposition~\ref{prop:lb} bound of \evMaxCombinedList{} groups, and \textsc{\acronym{}} fits all PPRs into \evMaxCombinedDepth{} group.

\subsection{Sensitivity}
Figure~\ref{fig:sensitivity} shows the sensitivity of each approach to the circuit parameters. The depth reductions decrease with the Pauli density (Figures~\ref{fig:sensitivity}a,~\ref{fig:sensitivity}h), with a Spearman correlation of $\evSpearmanDensity{}$, since dense PPRs anti-commute more often, forming smaller groups that rarely exceed the port budget. In contrast, they increase with the number of qubits and PPRs (Figures~\ref{fig:sensitivity}b,~\ref{fig:sensitivity}c,~\ref{fig:sensitivity}i,~\ref{fig:sensitivity}j), with correlations of \evSpearmanQubits{} and \evSpearmanPprs{}, respectively. Every additional pass improves the depth reduction of \textsc{\acronym{}} (Figures~\ref{fig:sensitivity}d,~\ref{fig:sensitivity}k), from \evGmPassesZeroT\% at $\ell = 0$ to \evGmPassesMaxT\% at $\ell = 9$ for Clifford\,+\,T and from \evGmPassesZeroRz\% to \evGmPassesMaxRz\% for Clifford\,+\,Rz, mostly in the first few passes.
We also vary the number of $X$ and $Z$ ports $c$, as allowed by other QEC codes or additional space overheads~\cite{cowtan_parallel_2026,zhang_time-efficient_2025}. The depth reductions of \textsc{\acronym{}} decrease as $c$ grows (Figures~\ref{fig:sensitivity}e,~\ref{fig:sensitivity}l), from \evGmPortsTwoT\% and \evGmPortsTwoRz\% at $c = 2$ to \evGmPortsMaxT\% and \evGmPortsMaxRz\% at $c = 16$, since anti-commutations more often end a group before the larger port budget is exceeded. For Clifford\,+\,T, this decrease comes almost entirely from \textsc{Restructure}, whose gain drops to \evGmPortsMaxRestructureT\% at $c = 16$, whereas \textsc{List} slightly improves from \evGmListT\% to \evGmPortsMaxListT\%. For Clifford\,+\,Rz, \textsc{List} also drops from \evGmListRz\% to \evGmPortsMaxListRz\%, and \textsc{Restructure}, whose gain falls to \evGmPortsMaxRestructureRz\%, accounts for the rest. However, \evNBumpSix{} circuits gain more at $c = 6$ than at $c = 2$. Finally, \textsc{\acronym{}} reduces the weight of the products by \evGmWeightCombinedT\% and \evGmWeightCombinedRz\% for Clifford\,+\,T and Clifford\,+\,Rz, respectively, at $c = 2$ (Figures~\ref{fig:sensitivity}f,~\ref{fig:sensitivity}g,~\ref{fig:sensitivity}m,~\ref{fig:sensitivity}n), a reduction that vanishes as $c$ grows. \textsc{List} thus remains promising even with better hardware, while \textsc{Restructure} is most effective under tight port budgets.
\section{Conclusion}
Fault-Tolerant Quantum Computation (FTQC) architectures permit parallel execution of commuting PPRs~\cite{litinski_game_2018,cowtan_parallel_2026}, but this parallelism is restricted by the number of hardware ``ports'' on each logical qubit, which increases the hardware-limited circuit depth. We formalize this problem, derive a lower bound on the hardware-limited depth, and show that computing it exactly is NP-hard. We propose two strategies to reduce this depth: port-aware list scheduling, which places each PPM in the earliest group that can accommodate it, and restructuring commuting PPMs into a generating set with lower port loads. Their combination, \acronym{}, reduces the hardware-limited depth by \evGmCombinedT\% and \evGmCombinedRz\% on average for Clifford\,+\, T and Clifford\,+\, Rz circuits, respectively, and by up to \evMaxCombined\%, and goes below the bound of Proposition~\ref{prop:lb} on \evBelowLBT{} and \evBelowLBRz{} circuits, respectively. Our analysis assumes unconstrained routing, and we leave routing-constrained architectures, such as planar surface code layouts, to future work.

\arxiv{
\section{Acknowledgements}
We would like to acknowledge Ted Yoder for insightful discussions related to executing commuting PPRs in parallel, and for suggesting us the idea of constructing equivalent generating sets in the context of commuting PPRs.

This work was funded in part by the Texas Quantum Institute (TQI).
}

\bibliographystyle{IEEEtranS}
\bibliography{references}

\end{document}